\documentclass[11pt]{article}
\title
{Unsatisfiable CNF-formulas}
\author{Heidi Gebauer
\thanks{Institute of
Theoretical Computer Science, ETH Zurich, CH-8092 Switzerland. Email:
gebauerh@inf.ethz.ch. } 
}

\usepackage{amsmath, amsfonts}
\usepackage{subfigure}
\usepackage[dvips]{graphicx}

\begin{document}
\bibliographystyle{plain}
\maketitle
\newtheorem{theo}{Theorem} [section]
\newtheorem{defi}[theo]{Definition}
\newtheorem{lemm}[theo]{Lemma}                                                                                                                                                                                                                                                                                                                                                                                                                                                                                                                   
\newtheorem{obse}[theo]{Observation}
\newtheorem{prop}[theo]{Proposition}
\newtheorem{coro}[theo]{Corollary}
\newtheorem{rem}[theo]{Remark}

\newcommand{\whp}{{\bf whp}}
\newcommand{\prob}{probability}
\newcommand{\rn}{random}
\newcommand{\rv}{random variable}
\newcommand{\hpg}{hypergraph}
\newcommand{\hpgs}{hypergraphs}
\newcommand{\subhpg}{subhypergraph}
\newcommand{\subhpgs}{subhypergraphs}
\newcommand{\bH}{{\bf H}}
\newcommand{\cH}{{\cal H}}
\newcommand{\cT}{{\cal T}}
\newcommand{\cF}{{\cal F}}
\newcommand{\cG}{{\cal G}}
\newcommand{\cD}{{\cal D}}
\newcommand{\cC}{{\cal C}}

\newcommand{\ideg}{\mathsf {ideg}}
\newcommand{\lv}{\mathsf {lv}}
\newcommand{\nga}{n_{\text{game}}}
\newcommand{\avdaneg}{\overline{\deg}}
\newcommand{\ed}{e_{\text{double}}}

\newcommand{\danger}{\mathsf {dang}}
\newcommand{\avdanan}{\overline{\danger}}

\newcommand{\degb}{\deg_{B}}
\newcommand{\degm}{\deg_{M}}

\newcommand{\avd}{\overline{D}}

\begin{abstract}
 A Boolean formula in a conjunctive normal form is called a \emph{$(k,s)$-formula} if every clause contains exactly $k$ variables and every variable occurs in at most $s$ clauses. We show that there are unsatisfiable $(k, 4 \cdot \frac{2^{k}}{k})$-CNF formulas.
\end{abstract}

\section{A better bound for unsatisfiable formulas}

\begin{theo} \label{theo:unsatisfiabilitylocal}
For every sufficiently large $k$ there is an unsatisfiable $(k, 4 \cdot \frac{2^{k}}{k})$-CNF.
\end{theo}
Note that due to Kratochv\'{i}l, Savick\'y and Tuza \cite{KST}  every $(k,\frac{2^{k}}{ek})$-CNF is satisfiable. So our result shows that this bound is tight up to a factor $4e$. 
\newline
\emph{Proof:} We consider the class $\cal{C}$ of hypergraphs $\cal{G}$ whose vertices can be arranged in a binary tree $T_{\cal{G}}$ such that every hyperedge of $\cal{G}$ is a path of $T_{\cal{G}}$. For positive integers $k,s \geq 1$ we denote by a \emph{$(k,s)$-tree} a $k$-uniform hypergraph  $\cal{G} \in \cal{C}$ such that 
\begin{itemize}
\item every full branch of $T_{\cal{G}}$ contains a hyperedge of $\cal{G}$ \enspace \enspace and
\item every vertex of $T_{\cal{G}}$ belongs to at most $s$ hyperedges of $\cal{G}$
\end{itemize}
When there is no danger of confusion we write $\cal{G}$ for $T_{\cal{G}}$.
The following lemma is the core of our proof.
\begin{lemm} \label{lemm:unsatisfiabilitylink}
For every sufficently large $k$ there is a $(k, 2 \cdot \frac{2^{k}}{k})$-tree $\cal{G}$.
\end{lemm}
We first show that Lemma \ref{lemm:unsatisfiabilitylink} implies Theorem \ref{theo:unsatisfiabilitylocal}.
Suppose that there is a $(k, 2 \cdot \frac{2^{k}}{k})$-tree $\cal{G}$ and let $\cal{G}'$ be a copy of $\cal{G}$. Let $\cal{H}$ be the hypergraph obtained by generating a new root $v$ and attaching $\cal{G}$ as a left subtree and $\cal{G}'$ as a right subtree.
Note that $\cal{H}$ is a $(k,2 \cdot \frac{2^{k}}{k})$-tree as well.

Let $(x_{1}, x'_{1}), (x_{2}, x'_{2}), \ldots, (x_{r}, x'_{r})$ denote the pairs of siblings of $\cal{H}$. We set $x'_{i} := \bar{x_{i}}$ for every $i$, $i = 1, \ldots, r$ (i.e. each non-root vertex represents a literal $x \in \{x_{1}, \bar{x_{1}}, x_{2}, \bar{x_{2}}, \ldots, x_{r}, \bar{x_{r}} \}$). Let $E(\cal{H})$ denote the set of hyperedges of $\cal{H}$. Then for every hyperedge $\{y_{1}, y_{2}, \ldots, y_{n}\} \in E(\cal{H})$ we form the clause $C_{\{y_{1}, y_{2}, \ldots, y_{n}\}} =  ({y_{1}} \vee {y_{2}} \vee \ldots \vee {y_{n}})$ and set ${\cal{F}} := \bigwedge_{ e \in E(\cal{H}) } C_{e}$.

Note that every variable $x_{i}$ of $\cal{F}$ occurs in at most $2 \cdot \Delta(F)$ clauses with $\Delta(F)$ denoting the maximum degree a variable in $\cal{F}$. Indeed, the number of occurrences of the variable $x_{i}$ is bounded by the number of occurrences of the literal $x_{i}$ plus the number of occurrences of the literal $\bar{x_{i}}$, which is at most $2 \Delta{F}$. So $\cal{F}$ is a $(k, 2 \cdot \frac{2^{k}}{k})$-CNF.

It remains to show that $\cal{F}$ is not satisfiable. Let $\alpha$ be an assignment to $\{x_{1}, \ldots, x_{r}\}$. 
\begin{obse} \label{obse:literalstrue}
Note that there is (at least) one full branch $b_{\text{full}}$ of $\cal{H}$ such that all literals along $b_{\text{full}}$ are set to FALSE by $\alpha$.
\end{obse}
By assumption $b_{\text{full}}$ contains a hyperedge $h$. But $\alpha$ does not satisfy the clause $C_{h}$, implying that $\alpha$ does not satisfy $\cal{F}$.
Since $\alpha$ was chosen arbitrarily, $\cal{F}$ is not satisfiable. $\Box$

It remains to prove our key lemma.
\newline
\emph{Proof of Lemma \ref{lemm:unsatisfiabilitylink}:} 
We need some notation first.
The vertex set and the hyperedge set of a {\hpg} $\cal{H}$ are denoted by $V(\cal{H})$ and $E(\cal{H})$, respectively. By a slight abuse of notation we consider $E(\cal{H})$ as a multiset, i.e. every hyperedge $e$ can have a multiplicity greater than 1.
By a \emph{bottom hyperedge} of a tree $T_{\cH}$ we denote a hyperedge covering a leaf of $T_{\cal{H}}$. 
Let $d = \frac{2^{k}}{k}$. For simplicity we assume that $k$ is a power of 2, implying that $d$ is power of 2 as well.

\begin{figure} [!htb]
\centering
\includegraphics[width=0.3\textwidth]{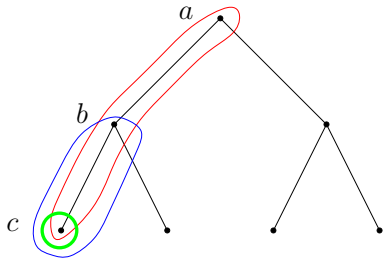}
\caption{An illustration of $\cal{H}$ for $d = 4$. The hyperedge $\{a,b,c\}$ has multiplicity 1,  $\{b,c\}$ has multiplicity 2 and $\{c\}$ has multiplicity 4.} \label{fig:WeakerClaim}
\end{figure}

To construct the required hypergraph $\cal{G}$ 
we establish first a (not necessarily $k$-uniform) hypergraph $\cal{H}$ and then successively modify its hyperedges and $T_{\cal{H}}$.
The following lemma is about the first step.
\begin{lemm} \label{lemm:First}
There is a hypergraph $\cal{H} \in \cal{C}$ with maximum degree $2d$ such that every full branch of $T_{\cal{H}}$ has $2^{i}$ bottom hyperedges of size $\log d + 1 - i$ for every $i$ with $0 \leq i \leq \log d$.
 \end{lemm} 
\emph{Proof of Lemma \ref{lemm:First}:} 
Let $T$ be a binary tree with $\log d + 1$ levels. In order to construct the desired hypergraph $\cal{H}$ we proceed for each vertex $v$ of $T$ as follows. 
For each leaf descendant $w$ of $v$ we let the path from $v$ to $w$ be a hyperedge of multiplicity $2^{l(v)}$ where $l(v)$ denotes the level of $v$.
Figure \ref{fig:WeakerClaim} shows an illustration.
The construction yields that each full branch of $T_{\cal{H}}$ has $2^{i}$ bottom hyperedges of size $\log d + 1 - i$ for every $i$ with $0 \leq i \leq \log d$. So it remains to show that $d(v) \leq 2d$ for every vertex of $v \in V(T)$. 
Note that every vertex $v$ has $2^{\log d - l(v)}$ leaf descendants in $T_{\cal{H}}$, implying that $v$ is the start node of  $2^{\log d - l(v)} \cdot 2^{l(v)} \leq d$ hyperedges. So the degree of the root is at most $d \leq 2d$. We then apply induction. Suppose that $d(u) \leq 2d$ for all nodes $u$ with $l(u) \leq i - 1$ for some $i$ with $1 \leq i \leq \log d$ and let $v$ be a vertex on level $i$. By construction exactly half of the hyperedges containing the ancestor of $v$ also contain $v$ itself. Hence $v$ occurs in at most $\frac{1}{2} \cdot 2d = d$ hyperedges as non-start node. Together with the fact that $v$ is the start node of at most $d$ hyperedges this implies that $d(v) \leq d + d \leq 2d$. $\Box$

The next lemma deals with the second step of the construction of the required hypergraph  $\cal{G}$.
\begin{lemm} \label{lemm:Second}
 There is a hypergraph $\cal{H'} \in \cal{C}$ with maximum degree $2d$ such that each full branch of $T_{\cal{H'}}$ has $2^{i}$ bottom hyperedges of size $\log d + 1 - i + \lfloor \log \log d \rfloor$ for some $i$ with $0 \leq i \leq \log d$.
 \end{lemm} 
\emph{Proof:} 
Let $\cal{H} \in \cal{C}$ be a hypergraph with maximum degree $2d$ such that every leaf $u$ of $T_{\cal{H}}$ is the end node of a set $S_{i}(u)$ of $2^{i}$ hyperedges of size $\log d + 1 - i$ for every $i$ with $0 \leq i \leq \log d$. (Lemma \ref{lemm:First} guarantees the existence of $\cal{H}$.) To each leaf $u$ of $T_{\cal{H}}$ we then attach a binary tree $T'_{u}$ of height $\lfloor \log \log d \rfloor$ in such a way that $u$ is the root of $T'_{u}$. 
Let $v_{0}, \ldots, v_{2^{\lfloor \log \log d \rfloor} - 1}$ denote the leaves of $T'_{u}$. For every $i$ with $0 \leq i \leq 2^{\lfloor \log \log d \rfloor} - 1$ we then augment every hyperedge of $S_{i}(u)$ with the set of vertices different from $u$ along the full branch of $T'_{u}$ ending at $v_{i}$.
 
After repeating this procedure for every leaf $u$ of $T_{\cal{H}}$ we get the desired hypergraph $\cal{H'}$. It remains to show that every vertex in $\cal{H'}$ has degree at most $2d$. To this end note first that during our construction the vertices of $\cal{H}$ did not change their degree. Secondly, let $u$ be a leaf of $T_{\cal{H}}$. By assumption $u$ has degree at most $2d$ and by construction $d(v) \leq d(u)$ for all vertices $v \in V(\cal{H'}) \backslash V(\cal{H})$, which completes our proof. $\Box$

\begin{lemm} \label{lemm:NThird}
There is a hypergraph $\cal{H''} \in \cal{C}$ with maximum degree $2d$ such that every full branch of $T_{\cal{H''}}$ has one bottom hyperedge of size $\log d + 1  + \lfloor \log \log d \rfloor$.
 \end{lemm} 
Note that due to our choice of $d$, Lemma \ref{lemm:NThird} directly implies Lemma \ref{lemm:unsatisfiabilitylink}. $\Box$
 \newline
 \emph{Proof of Lemma \ref{lemm:NThird}:} By Lemma \ref{lemm:Second} there is a hypergraph $\cal{H'} \in \cal{C}$ with maximum degree $2d$ such that each full branch of $T_{\cal{H'}}$ has $2^{i}$ bottom hyperedges of size $\log d + 1 - i + \lfloor \log \log d \rfloor$ for some $i$ with $0 \leq i \leq \log d$. For every leaf $u$ of $T_{\cal{H'}}$ we proceed as follows. Let $e_{1}, \ldots, e_{2^{i}}$ denote the bottom hyperedges of $\cal{H'}$ ending at $u$. We then attach a binary tree $T''$ of height $i$ to $u$ in such a way that $u$ is the root of $T''$. 
Let $p_{1}, \ldots, p_{2^{i}}$ denote the full branches of $T''$. We finally augment $e_{j}$ with the vertices along $p_{j}$, for $j = 1 \ldots 2 ^{i}$.

After repeating this procedure for every leaf $u$ of $T_{\cal{H'}}$ we get the resulting graph $\cal{H''}$.
By construction every full path of $T_{\cal{H''}}$ has one bottom hyperedge of size $\log d + 1  + \lfloor \log \log d \rfloor$.
A similar argument as in the proof of Lemma \ref{lemm:Second} shows that the maximum degree of $\cal{H''}$ is at most $2d$. $\Box$


\end{document}